\RequirePackage{luatex85}
\documentclass[a4paper,11pt]{article}
\usepackage{pos}
\usepackage[qm]{qcircuit}
\usepackage{mathtools}
\usepackage{braket}
\usepackage{ctable}
\usepackage{stmaryrd}

\title{Noisy Bayesian optimization for variational quantum eigensolvers}

\author*[a,b,c,d]{Giovanni Iannelli}
\author[a]{Karl Jansen}

\affiliation[a]{Deutsches Elektronen-Synchrotron DESY,\\
  Platanenallee 6, 15738 Zeuthen, Germany}

\affiliation[b]{Department of Physics, University of Cyprus,\\
Panepistimiou Street 1, 2109 Aglantzia, Nicosia, Cyprus}

\affiliation[c]{Institut für Physik, Humboldt-Universität zu Berlin,\\
Newtonstra{\ss}e 15, 12489 Berlin, Germany}

\affiliation[d]{Dipartimento di Fisica, Università degli Studi di Roma ``Tor Vergata'',\\
Via della Ricerca Scientifica 1, 00133 Rome, Italy}

\emailAdd{giovanni.iannelli@desy.de}
\emailAdd{karl.jansen@desy.de}

\abstract{The variational quantum eigensolver (VQE) is a hybrid quantum-classical algorithm
used to find the ground state of a Hamiltonian using  variational methods.
In the context of this Lattice symposium, the procedure can be used to study lattice gauge theories (LGTs) in the Hamiltonian formulation.
Bayesian optimization (BO) based on Gaussian process regression (GPR)
is a powerful algorithm for finding the global minimum of a cost function, e.g. the energy, with a very low number 
of iterations using data affected by statistical noise.
This work proposes an implementation of GPR and BO specifically tailored to perform VQE on quantum computers already available today.}

\FullConference{%
 The 38th International Symposium on Lattice Field Theory, LATTICE2021
  26th-30th July, 2021
  Zoom/Gather@Massachusetts Institute of Technology
}

\begin{document}
\maketitle

\section{Introduction}

Quantum algorithms have the potential to be exponentially quicker than classical alternatives in many noteworthy scientific applications.
Examples are quantum machine learning \cite{biamonte2017quantum}, quantum chemistry \cite{lanyon2010towards},
and many others \cite{montanaro2016quantum}. Unfortunately, many of these applications are not yet implementable on current noisy intermediate-scale quantum (NISQ)
computers \cite{preskill2018quantum} and need to wait until noise sources can be suppressed down
to a threshold that makes quantum computers usable in practice or to even build fault-tolerant quantum computers \cite{shor1996fault}.

However, many interesting problems of LGTs can already be studied with NISQ devices \cite{banuls2020simulating}. In particular, if LGTs are studied in their Hamiltonian formulation, quantum algorithms do not generally suffer from the sign problem \cite{kan2021investigating,zhang2021simulating}.
An important ready-to-use algorithm is the variational quantum eigensolver (VQE) \cite{peruzzo2014variational},
which is a hybrid quantum-classical algorithm for finding the ground (and excited) state of a given Hamiltonian $\mathcal H$
using the variational principle.
The \emph{quantum part} of VQE deals with measuring the expectation value of the Hamiltonian, i.e. the energy, in a given multi-qubit state,
while the \emph{classical part} consists of searching among a family of multi-qubit states generated by a parametrized quantum circuit
to find the state that minimizes the energy.

The algorithm proposed in this proceedings is a classical optimizer that aims to find a good approximation of the ground state reducing
as much as possible the number of energy measurements.
The approach chosen here is known as Bayesian global optimization. Its first application dates back in the 60s \cite{kushner1964new},
while its modern implementations are based on a more recent work \cite{jones1998efficient}.
The backbone of this method is Gaussian process regression (GPR),
which is an interpolation method based on Bayesian inference of Gaussian processes.
It allows us to create predictive models of black-box functions using a limited amount of (noisy) data.
At each optimization iteration, this model is used to determine a set of parameters presumably close to the global minimum point.
This step is performed following a procedure called acquisition function optimization.

The algorithm proposed here to optimize the energy differs from the other alternatives commonly used in VQE
as it uses not only the estimated values of the energy, but also the values of their statistical errors.
The motivation is to lower the number of quantum measurements at each step:
the procedure is well defined even for imprecise energy measurements as long as their errors is approximately Gaussian
due to the central limit theorem.
Results of this algorithm are compared to other commonly chosen alternatives using simulators of noisy devices.

\section{Quantum expectation estimation}

Given a Hamiltonian $\mathcal H$, it first needs to be written as a polynomial of sigma matrices:
\begin{equation}\label{eq:hamiltonian}
    H = \sum_{i\gamma}h_\gamma^i\sigma_\gamma^i+
    \sum_{ij\gamma\delta}h_{\gamma\delta}^{ij}\sigma_\gamma^i\otimes\sigma_\delta^j+\ldots
\end{equation}
where the $h$s are real parameters, Latin indices identify the qubit on which sigma matrices are acting, and Greek indices the sigma matrices coordinates. The quantum expectation estimation (QEE) algorithm \cite{peruzzo2014variational} computes the expectation value of the energy
$E_\psi \equiv \Braket{\psi|\mathcal H|\psi}$ for any input multi-qubit state $\ket \psi$
with a possible quantum advantage with respect to equivalent classical approaches.
Furthermore, QEE is already implementable in NISQ devices as the computation of $E_\psi$
can be decomposed in many short quantum programs, therefore reducing the impact of quantum noise.

However, due to the probabilistic nature of quantum measurements,
we have only access to a stochastic variable that estimates $E_\psi$
(see for example \cite{kandala2017hardware} for the Qiskit implementation).
In order to get a precise estimation, it is possible to 
perform multiple independent measurements, also called \emph{shots}, and use their mean as an estimator of $E_\psi$. Let us consider the case of measurements in the absence of quantum noise. As the number of shots goes to infinity, the central limit theorem tells us that 
the mean converges to a Gaussian centered in $E_\psi$ whose variance is estimated by the standard error of the mean.
Therefore, given a number $S$ of independent shots, calling $E^1_\psi, ..., E^S_\psi$
the energy measurements of each single shot, it is possible 
to measure the following energy estimator on a quantum computer:
\begin{gather}
    \hat E^S_\psi \equiv  \frac{1}{S}\sum_{m=1}^S E^m_\psi \label{eq:estimator}\\
    \text{Var}[\hat E^S_\psi] = \frac{1}{S(S-1)}\sum_{m=1}^S(E^m_\psi-\hat E^S_\psi)^2 \label{eq:error}
\end{gather}

It is important to emphasize that there is a difference between the \emph{statistical noise} and the \emph{quantum noise}. The statistical noise is the (approximately) Gaussian deviation of $\hat E^S_\psi$ caused by the probabilistic nature of quantum measurements, while quantum noise is the deviation caused by the imperfections of real quantum devices.
The impact of quantum noise to QEE is to add a BIAS to the estimator of Eq.~\eqref{eq:estimator}.
This BIAS can be significantly reduced using error mitigation techniques (a comparison can, e.g., be found in Section 5.1 of \cite{funcke2020measurement}).

In the rest of these proceedings, for simplifying the notation, the number of shots $S$ will be omitted. 
Since on the quantum computer a parametrized quantum circuit is employed with parameters $\theta_\alpha$, we 
will denote the parametrized energy $E(\theta_\alpha)$ with error $\Delta E(\theta_\alpha)$ with 
$E(\theta_\alpha) \equiv \hat E^S(\theta_\alpha)$
and $\Delta E(\theta_\alpha)\equiv\sqrt{\text{Var}[\hat E^S(\theta_\alpha)]}$.

\section{Variational quantum eigensolver}

The objective Hamiltonian $\mathcal H$ needs first to be written in the form of, e.g., Eq.~\eqref{eq:hamiltonian}.
Then, a family of qubit states $\ket{\psi(\theta_\alpha)}$ is introduced with a $d$-dimensional parameter set $\theta_\alpha$.
This can be achieved by applying a parametrized quantum circuit $U(\theta_\alpha)$ on a fixed initial multi-qubit state,
usually chosen to be $\ket{0 \cdots 0}$:
\begin{equation}
    \ket{\psi(\theta_\alpha)} \equiv U(\theta_\alpha)\ket{0 \cdots 0}\label{eq:parametrization}
\end{equation}
This state parametrization allows us to define a parametrized energy:
\begin{equation}
    E(\theta_\alpha)\equiv\Braket{\psi(\theta_\alpha)|\mathcal H|\psi(\theta_\alpha)}\label{eq:param_energy}\\
\end{equation}
which can be evaluated using the QEE algorithm for any value of $\theta_\alpha$.

Then, the VQE consists of approximating the ground state $\ket{\psi_{\min}}$ performing the following optimization:
\begin{equation}\label{eq:optimization}
    \text{min}_{\theta_\alpha}E(\theta_\alpha)
    \geq \Braket{\psi_{\min}|\mathcal H|\psi_{\min}} = E_{\min}
\end{equation}

It is important to note that most of the optimizers commonly used within VQE take as input only measurements of the estimator in Eq.~\eqref{eq:estimator}
and not of its error in Eq.~\eqref{eq:error}.
In some cases they rely on performing energy measurements that are precise enough to be considered
(almost) exact, which means choosing the number of shots $S$ to be large enough so that the error in Eq.~\eqref{eq:error}
can be neglected.
This is done, for example, in the original VQE paper \cite{peruzzo2014variational}
using the Nelder-Mead \cite{nelder1965simplex} optimizer,
as well as in a well-performing recently published optimizer algorithm \cite{nakanishi2020sequential}
specifically tailored for VQE.
On the other hand, some algorithms leverage on the statistical noise of input measurements to escape local minima, as for example in the SPSA \cite{spall1992multivariate} optimizer.
The algorithm proposed in this proceedings differs from the considered alternatives
as it uses both the estimator of Eq.~\eqref{eq:estimator}
and its error of Eq.~\eqref{eq:error}.

\section{Bayesian optimization}

In the Bayesian optimization approach one first needs an initialization step with a sequence of energy measurements obtained with circuit parameters chosen (quasi) randomly.
After this initial step, the core of the algorithm consists of two building blocks. At each iteration, it first uses GPR to create a predictive model of the parametrized energy in Eq.~\eqref{eq:param_energy} using the information of the previous energy measurements.
Then, this predictive model is used to define an acquisition function that assigns to each circuit parameter values a positive score. The parameters that maximize the acquisition function, i.e. the one with the highest score, will then be chosen for the next energy measurement.

\subsection{Gaussian process regression}

A Gaussian process (GP) $f$ maps a set of $n$ $d$-dimensional circuit parameters $\theta_{1\alpha}, ..., \theta_{n\alpha}$
to $n$ stochastic variables $f_1, ..., f_n$ distributed as: 
\[
    p(f_i) = \det(2\pi K)^{-\frac{1}{2}}\exp\left(-\frac{1}{2}\sum_{ij}(f_i-\mu_i)(K^{-1})_{ij}(f_j-\mu_j)\right)
\]
where $\mu_i\equiv\mu(\theta_{i\alpha}), K_{ij}\equiv k(\theta_{i\alpha},\theta_{j\alpha}')$, $\mu(\theta_\alpha)$ is the GP mean function and $k(\theta_{\alpha},\theta_{\alpha}')$ is the GP covariance function.

Given a set of energy measurements $E_1, ..., E_n$ obtained with the corresponding circuit parameter values
$\theta_{1\alpha}, ..., \theta_{n\alpha}$ with their respective measurement errors $\Delta E_1, ..., \Delta E_n$,
Gaussian process regression\footnote{Also known as kriging in geostatistics.} (GPR) \cite{matheron1962traite}
is a procedure that finds a GP whose mean function $\mu(\theta_\alpha)$ interpolates the unknown energy functions $E(\theta)$. For details about GPR see, e.g., the textbook \cite{williams2006gaussian}. For what concerns our presentation, we report the analytical result for the mean and covariance functions of the posterior obtained with GPR:
\begin{equation}\label{eq:gpr}\begin{dcases}
	\mu(\theta_\alpha|E_i) = \mu(\theta_\alpha)+
		\sum_{ij}k(\theta_\alpha,\theta_{i\alpha})(\widetilde K^{-1})_{ij}(E_j-\mu(\theta_{j\alpha})) \\
	k(\theta_\alpha,\theta'_\alpha|E_i) = k(\theta_\alpha,\theta'_\alpha)-
		\sum_{ij}k(\theta_\alpha,\theta_{i\alpha})(\widetilde K^{-1})_{ij}k(\theta_{j\alpha},\theta_\alpha')
\end{dcases}\end{equation}
where $\widetilde K_{ij} \equiv k(\theta_{i\alpha},\theta_{j\alpha}) + \Delta E_i^2 \delta_{ij}$. The posterior mean $\mu(\theta_\alpha|E_i)$ is our surrogate model of the unknown parametrized energy, while $k(\theta_\alpha,\theta_\alpha'|E_i)$ is an estimation of its Gaussian covariance. We emphasize that the measurement errors evaluated with Eq.~\eqref{eq:error} are used in the evaluation of $\widetilde K_{ij}$, and this formula is exact in the case of Gaussian errors, which is asymptotically true as the number of shots grows.

The choice of the prior mean $\mu(\theta)$ and the prior covariance function $k(\theta_\alpha,\theta_\alpha')$ is subjective, as it usually happens in the context of Bayesian inference. Their choice has an impact on the geometry of the posterior GP of Eq.~\eqref{eq:gpr}.
The possibility of using different $\mu(\theta)$ and $k(\theta_\alpha,\theta_\alpha')$ can then be an advantage as it can be used to impose certain properties that are motivated from the physics of the considered problem.
For example, in the noiseless case, the parametrized energy $E(\theta_\alpha)$ of Eq.~\eqref{eq:param_energy} is in $C^\infty$ when using commonly chosen quantum circuits \cite{nakanishi2020sequential}, and selecting $\mu(\theta_\alpha),k(\theta_\alpha,\theta_\alpha')\in C^\infty$ will impose this property on the posterior mean of Eq.~\eqref{eq:gpr}.
A common choice is to set $\mu(\theta_\alpha)$ to a constant:
\begin{equation}\label{eq:mean}
    \mu(\theta_\alpha) = \mu
\end{equation}and $k(\theta_\alpha,\theta_\alpha')$ to the RBF kernel:
\begin{equation}\label{eq:rbf}
k^\text{RBF}(\theta_\alpha,\theta_\alpha')=\sigma^2\prod_\alpha \exp\left(-\frac{(\theta_\alpha-\theta_\alpha')^2}{2\ell_\alpha^2}\right)
\end{equation}
where $\mu$, $\sigma$, $\ell_\alpha$ are hyperparameters that can be fixed with maximum likelihood estimation of type II
(MLE-II) \cite{williams2006gaussian}.

In most applications, the energy $E(\theta_\alpha)$ of Eq.~\eqref{eq:param_energy} is not only $C^\infty$, but $2\pi$-periodic for each $\alpha=1, ..., d$. This property can be imposed to a GP using as covariance function the periodic kernel \cite{mackay1998introduction}:
\begin{equation}\label{eq:periodic}
k^P(\theta_\alpha,\theta_\alpha') = \sigma^2\prod_\alpha\exp\left(-\frac{2}{\ell_\alpha^2}\sin^2\left(\frac{\theta_\alpha-\theta_\alpha'}{2}\right)\right)
\end{equation}

\subsection{Acquisition function}
The circuit parameter values of each optimization step are chosen to be the maximum of an appropriately defined acquisition function.
A common choice is the expected improvement (EI) acquisition function \cite{movckus1975bayesian}. Calling $E_\text{min}\equiv\min(E_{1\alpha}, ..., E_{n\alpha})$ the minimum of the previously measured energies and $\hat E(\theta_\alpha)$ the energy prediction given by the surrogate model, the EI is defined as:
\begin{equation}\label{eq:ei}
a_\text{EI}(\theta_\alpha) \equiv \mathbb E_{\hat E(\theta_\alpha)}[\max(0,E_\text{min}-\hat E(\theta_\alpha))]
\end{equation}

where the expectation value $\mathbb E_{\hat E(\theta_\alpha)}[...]$ is evaluated among all the possible values of $\hat E(\theta_\alpha)$ evaluated from the surrogate model.

A great feature of the EI is that both $a_\text{EI}(\theta_\alpha)$ and its gradient $\partial_\alpha a_\text{EI}(\theta_\alpha)$ are available in closed form if a GP is used as surrogate model \cite{jones1998efficient}. While the EI proves to be very effective for noiseless objective functions, it is not as effective in presence of statistical noise \cite{vazquez2008global}. Since this is the case for VQE, a better choice is an extension of the EI called noisy expected improvement (NEI) \cite{letham2019constrained}.

The NEI is an extension of the EI that is not available in closed form. However, it is possible to efficiently evaluate it with (quasi-)Monte Carlo methods. Let $\mathcal E_1, ...,\mathcal E_K$ be noiseless (quasi)random energy functions sampled from the posterior of Eq.~\eqref{eq:gpr}, and let $a_\text{EI}(\theta_\alpha|\mathcal E_1), ..., a_\text{EI}(\theta_\alpha|\mathcal E_K)$ be EIs defined over them. Then:
\begin{equation}\label{eq:approx_nei}
a_\text{NEI}(\theta_\alpha) \simeq \frac{1}{K}\sum_i a_\text{EI}(\theta_\alpha|\mathcal E_i)
\end{equation}
Each summand of Eq.~\eqref{eq:approx_nei} is available in closed form so that its evaluation can be done fast. Furthermore, also its gradient is computable analytically, which can be useful for optimizing the acquisition function.

\section{Outline of the algorithm}
Here we describe our proposed algorithm step by step and specify details of its implementation.
The Hamiltonian and the parametrized circuit of Eq.~\eqref{eq:parametrization} are problem dependent. Quantum computing libraries have procedures for evaluating the parametrized energy of Eq.~\eqref{eq:param_energy} with any value of the $d$-dimensional circuit parameter $\theta_\alpha$ using the estimator in Eq.~\eqref{eq:estimator} and its error of Eq.~\eqref{eq:error}. This measurement is obtained performing $S$ shots. We used the Qiskit \cite{aleksandrowicz2019qiskit} quantum computing library for our tests.

Once the routine for measuring $E(\theta_\alpha)$ is defined, the Bayesian optimization procedure is entirely implemented on a classical computer. We built our test on top of the libraries Ax \cite{bakshy2018ae}, BoTorch \cite{balandat2020botorch} and GPyTorch \cite{gardner2018gpytorch}. The Bayesian optimization is then performed as follows:
\begin{enumerate}
\item
    Generate $n$ quasi-random $d$ dimensional points $\theta_{1\alpha}, ..., \theta_{n\alpha} \in [0,2\pi]^d$ with a Sobol sequence \cite{sobol1967distribution}. In our tests, we used $n=3$.
\item
    Given $\theta_{1\alpha}, ..., \theta_{n\alpha}$, measure their corresponding energies $E_1, ..., E_n$ and their errors $\Delta E_1, ..., \Delta E_n$ as described in Eq.~\eqref{eq:estimator} and Eq.~\eqref{eq:error}.
\item
    Use MLE-II to infer the prior hyperparameter $\mu$ of Eq.~\eqref{eq:mean} and $\sigma$, $\ell_\alpha$ of Eq.~\eqref{eq:rbf} or Eq.~\eqref{eq:periodic}, depending on whether RBF or periodic kernel was chosen.
    The default settings of Ax, BoTorch and GPyTorch were used for this inference.
\item
    Compute the GP posterior mean and covariance of Eq.~\eqref{eq:gpr}.
\item
    The current estimation of the parameters for the global minimum point $\theta_\alpha^\text{min}$ is chosen among $\theta_{1\alpha}, ...,\theta_{n\alpha}$ as the $\theta_{i\alpha}$ with the minimum expected energy according to the GP model found at point 4:
    \begin{equation}\label{eq:minimum}
    \theta_\alpha^\text{min}\equiv \text{argmin}_{\theta_{i\alpha}}\ \mu(\theta_{i\alpha}|E_i)
    \end{equation}
    and the corresponding estimation of the minimum energy $E^\text{min}$ is:
    \begin{equation}\label{eq:min_energy}
    E^\text{min}\equiv \mu(\theta_\alpha^\text{min}|E_i)
    \end{equation}
\item
    Sample $K$ noiseless energy functions from the posterior GP found at point 4.
    For each of the $K$ samples, perform a different noiseless GPR using the same hyperparameters found in point 3, and compute the $K$ EIs necessary for the NEI approximation of Eq.~\eqref{eq:approx_nei}. In our tests, we used $K=20$.
\item
    Perform a global optimization of the approximated NEI to find its maximum point $\theta_{\alpha}^\text{NEI}$. We performed this optimization with the default procedure of BoTorch, which is a multistart L-BFGS-B \cite{byrd1995limited}, where 20 restart points are selected as those with the maximum acquisition function value out of 1000 points drawn from the Sobol sequence in $[0,2\pi]^d$. The SciPy \cite{virtanen2020scipy} implementation of L-BFGS-B is used for this step.
\item
    Add the NEI maximum found at point 7 to the parameter set $\theta_{n+1,\alpha}\,\mapsfrom\,\theta_{\alpha}^\text{NEI}$ and iterate from point 2 with $n\,\mapsfrom\,n+1$ until a break condition is reached which could be realized, e.g., when the minimum of Eq.~\eqref{eq:minimum} is stable for a certain number of iterations. Alternatively a fixed number $N$ of iterations might be chosen beforehand in order to keep the number of quantum measurements under control.
\end{enumerate}

\section{Testing with IBMQ}

We tested the algorithm described in section 5 on a simple two qubits Hamiltonian of the  transverse-field Ising model with coupling set to one:
\begin{equation}\label{eq:ising}
    \mathcal H = -\sigma_x^1\otimes\sigma_x^2-\sigma_z^1-\sigma_z^2
\end{equation}

The parametrization of Eq.~\eqref{eq:parametrization} was achieved using the following quantum circuit:
\begin{equation*}
    \Qcircuit @C=.75em @R=.5em @!R {
        \lstick{ \ket 0  } & \gate{R_y({\ensuremath{\theta_1}})} & \ctrl{1} & \gate{R_y({\ensuremath{\theta_3}})} & \gate{R_z({\ensuremath{\theta_{5}}})} & \qw\\
    \lstick{ \ket 0  } & \gate{R_y({\ensuremath{\theta_2}})} & \targ & \gate{R_y({\ensuremath{\theta_4}})} & \gate{R_z({\ensuremath{\theta_6}})} & \qw\\
    }
\end{equation*}
This circuit was constructed using the procedure described in \cite{funcke2021dimensional}. It does not have redundant parameters and can cover the whole Hilbert space if we exclude states which are equivalent after the application of a global phase.

For the quantum measurement, we used the Qiskit simulator using the noise model of IBMQ Santiago quantum device. Assuming that only a fixed total number of shots is at our disposal, we tried different algorithms with different number of shots per measurement in order to find the setup that uses this assumed budget most efficiently.
We first analyze the results obtained with two algorithms that have been proposed for this specific task: the SPSA \cite{spall1992multivariate} in its implementation \cite{kandala2017hardware} available in Qiskit, and the NFT \cite{nakanishi2020sequential}\footnote{Its implementation is available at \url{https://github.com/ken-nakanishi/nftopt}.}. Then we compare their performance with two implementations of BO using respectively the RBF and the periodic kernels of Eqs.~\eqref{eq:rbf} and \eqref{eq:periodic}. Each algorithm is tested using 20, 40 and 80 number of measurement, each respectively obtained with 64, 32 and 16 number of shots. Therefore, the total number of shots used is 1280 in all cases.

The optimization results are then compared to the exact values of the ground state and the ground state energy. In particular, it is possible to evaluate the state fidelity of a parametrized state with respect to the exact ground state. The fidelity is equal to the square of the scalar product of two states and it quantifies their proximity as its maximum value of 1 is reached when the two states are identical.

After each iteration of the tested algorithms, both the energy and the fidelity have been recorded. The results shown are the average of what was obtained with 20 independent runs starting with different random initial conditions. Their error corresponds to the standard error computed out of these 20 repetitions.

\begin{figure}[!htb]
    \centering
    \includegraphics{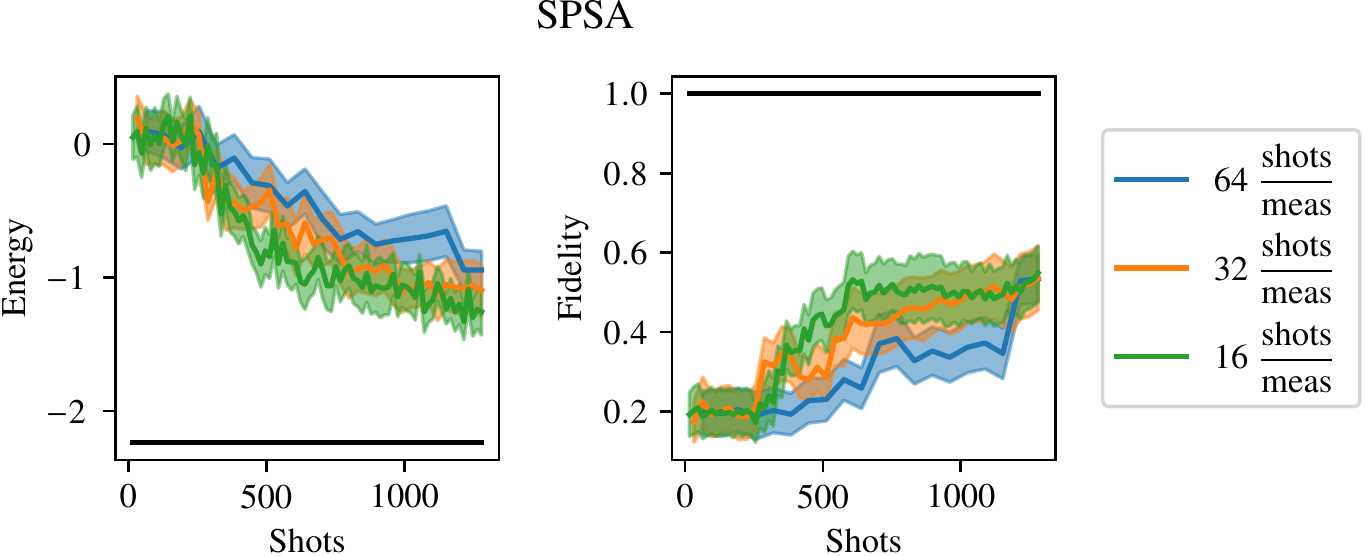}
    \caption{shots/meas indicates the number of shots used for each energy measurement. Black lines correspond to the exact solution.}
    \label{fig:spsa}
\end{figure}

\paragraph{SPSA}
The results for SPSA\footnote{The Qiskit implementation of SPSA was used with its default settings.} are shown in Fig.~\ref{fig:spsa}. The performances slightly improve decreasing the number of shots per measurements, but none of the three setups gets close to the exact solution.

\begin{figure}[!htb]
    \centering
    \includegraphics{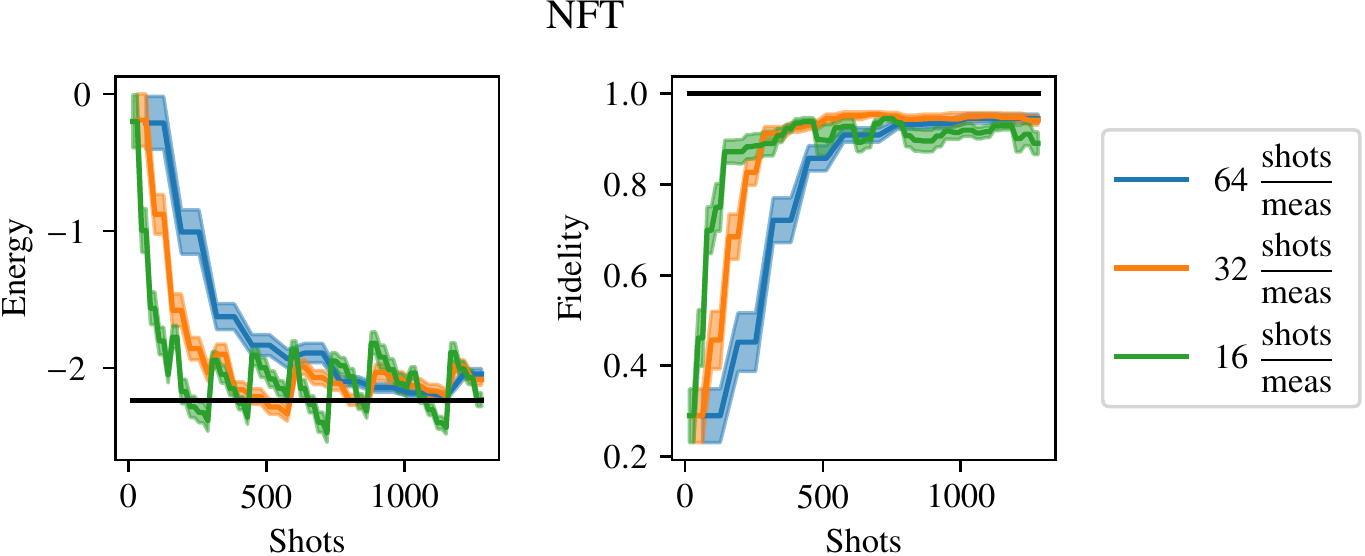}
    \caption{Here is used the same notation of Fig.~\ref{fig:spsa}}
    \label{fig:nft}
\end{figure}

\paragraph{NFT}
The performances of the NFT algorithm\footnote{These results were obtained setting the variable {\ttfamily reset\_interval=4} in the function made available by the authors.} are shown in Fig.~\ref{fig:nft}. In this case, the optimization gets quickly close to the solution in all three cases. The speed of convergence increases reducing the number of shots per measurement, but, on the other hand, the stability and the precision of the solution decreases. The fidelity is slightly unstable only with 16 shots per measurement, while energy measurements do not have a good precision with 16 and 32 shots.

The NFT algorithm has overall a very quick convergence rate and a low requirement of CPU resources due to its usage of an analytical formula of the target energy. However, its solutions do not have a good precision with very small number of shots per measurements as this algorithm does not have built-in methods to infer the real value of the energy, but it has to rely on the mean value of Eq.~\eqref{eq:estimator}.

\begin{figure}[!htb]
    \centering
    \includegraphics{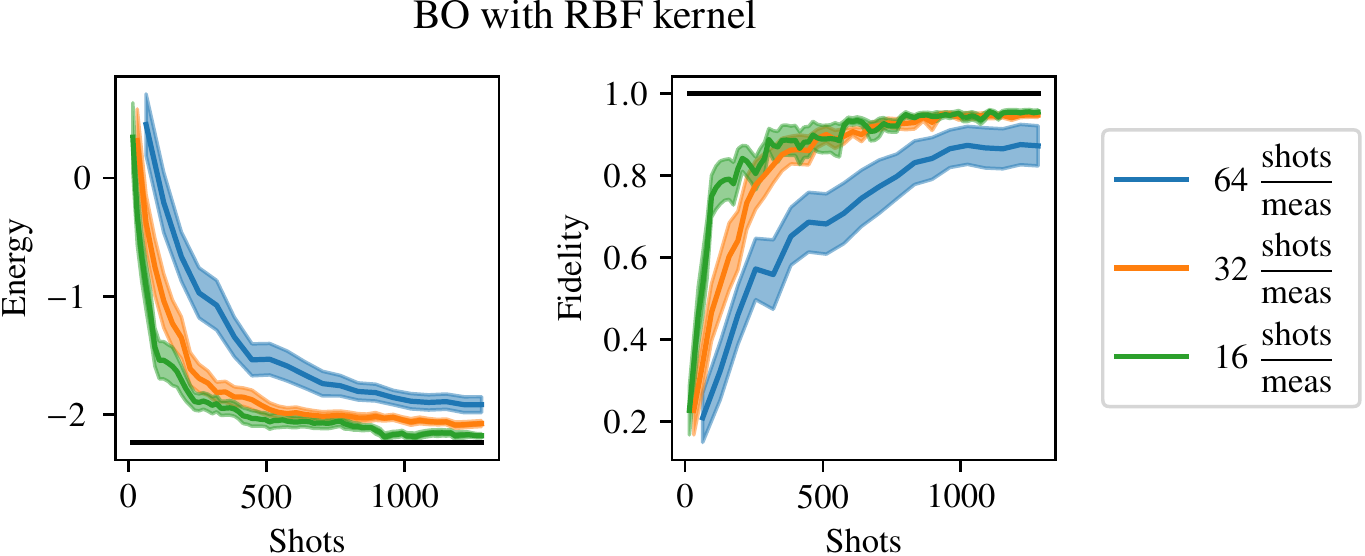}
    \caption{Here is used the same notation of Fig.~\ref{fig:spsa}}
    \label{fig:rbf}
\end{figure}

\begin{figure}[!htb]
    \centering
    \includegraphics{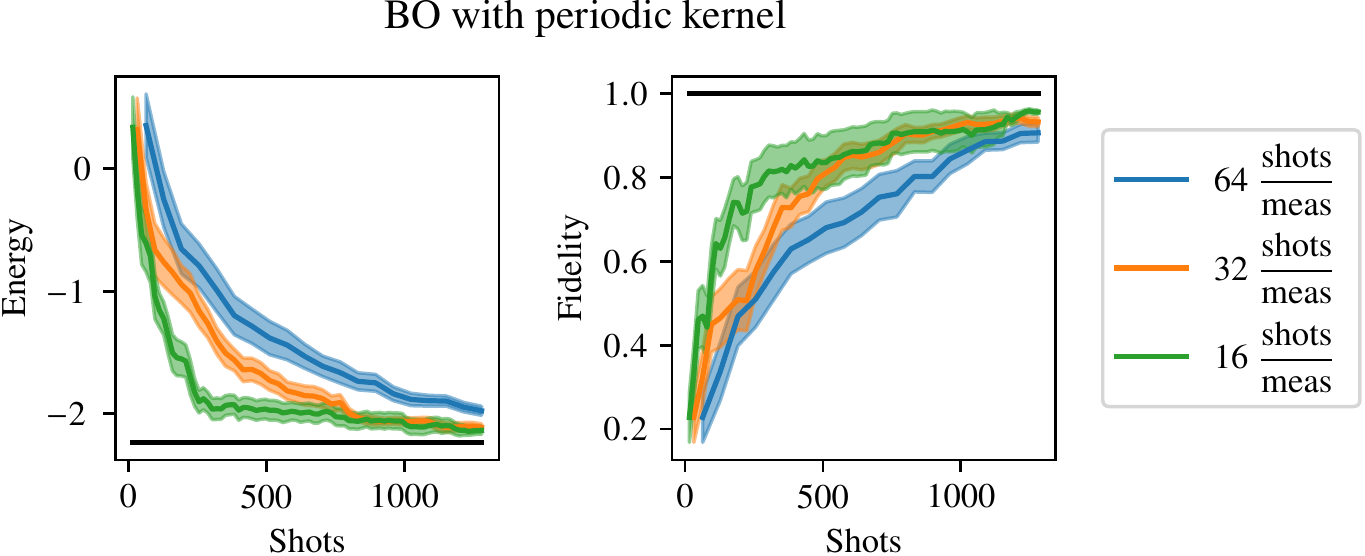}
    \caption{Here is used the same notation of Fig.~\ref{fig:spsa}}
    \label{fig:periodic}
\end{figure}

\paragraph{Bayesian optimizer}
Results obtained with the RBF kernel are reported in Fig.~\ref{fig:rbf}, while in Fig.~\ref{fig:periodic} we show those obtained with the periodic kernel. The two implementations behave in a very similar way. In both cases the performances improve reducing the number of shots per measurement and the solution gets closer to the exact values without a loss of stability.

\begin{figure}[!htb]
    \centering
    \includegraphics{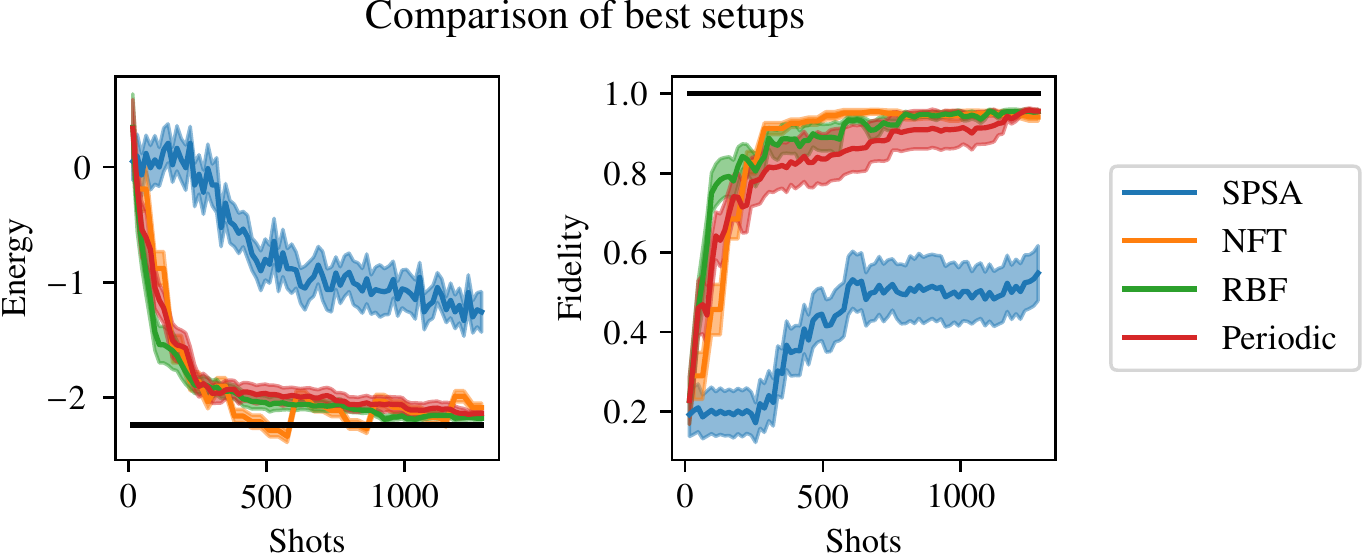}
    \caption{Here are compared the best results of each previous plot.}
    \label{fig:comparison}
\end{figure}

\paragraph{Comparison of the algorithms}
In Fig.~\ref{fig:comparison} we report the results of each of the considered algorithms in their best setup, which is 32 shots per measure for NFT and 16 for the others. SPSA is outperformed by NFT and the BOs, which have similar performances in this setup. However, the BOs have better estimation of the ground state energy as its value is inferred with GPR. It removes the tradeoff present in NFT between precision and speed at the cost of increasing CPU time. RBF initially converges faster than periodic, but at the end the solution with the highest fidelity was found with the periodic kernel.

\section{Conclusions and outlooks}
Our conclusions are that BO using GPR and NEI is a good choice for VQE in case of noisy measurements obtained with a few number of shots as it can use both the energy mean value of Eq.~\eqref{eq:estimator} and its error of Eq.~\eqref{eq:error}. It outperforms SPSA in the here considered case and has a convergence rate similar to the one obtained with NFT. However, BO provides a more precise estimation of the ground state energy, although at the cost of increasing the CPU time, which is presently clearly not a bottleneck when compared to the available QPU time. BO with RBF kernel started converging faster than with periodic kernel, but the most accurate solution was found with the periodic kernel.

The main weakness of BO is its expansion to a high number of circuit parameters, at its standard implementation would require too much CPU time and memory. With some modifications, BO was, however, successfully used with a high number of parameters in other contexts (see for example \cite{wang2016bayesian,li2017high,kirschner2019adaptive}). At the moment we are exploring different possible ways to expand the BO of VQEs to cases in which the number of circuit parameters grows up to $\mathcal O(100)$, which 
would provide the BO approach a promising perspective to be used in VQE also for the next generation of quantum computers. 

\section{Aknowledgements}
This project has received funding from the Marie Skłodowska-Curie European Joint Doctorate program STIMULATE of the European Commission under grant agreement No.~765048. G.I.'s position was funded under this program.

\clearpage
\bibliographystyle{JHEP}
\bibliography{bibliography}

\end{document}